\documentclass[pra,aps,showpacs,preprint]{revtex4}
\usepackage{graphicx}
\usepackage{amsmath}

\begin{document}
\title{\bf Quantum correlation evolution of GHZ and W
states under noisy channels using ameliorated measurement-induced
disturbance}
\author{Pakhshan Espoukeh}
\author{Pouria Pedram}
\email[Electronic address: ]{p.pedram@srbiau.ac.ir}
\affiliation{Department of Physics, Science and Research Branch,
Islamic Azad University, Tehran, Iran}
\date{\today}

\begin{abstract}
We study quantum correlation of Greenberger-Horne-Zeilinger (GHZ)
and W states under various noisy channels using measurement-induced
disturbance approach and its optimized version. Although these
inequivalent maximal entangled states represent the same quantum
correlation in the absence of noise, it is shown that the W state is
more robust than the GHZ state through most noisy channels. Also,
using measurement-induced disturbance measure, we obtain the
analytical relations for the time evolution of quantum correlations
in terms of the noisy parameter $\kappa$ and remove its
overestimating quantum correlations upon implementing the
ameliorated measurement-induced disturbance.
\end{abstract}

 \maketitle

\section{Introduction}
Quantification of correlations in bipartite quantum systems is one
of the important problems in quantum information. Quantum
correlations can be considered as the resources for quantum
information processes \cite{nel}. Initially, it was assumed that
entanglement which plays an important role in quantum computing and
quantum information processing, is the only kind of nonclassical
correlation in a quantum state. This issue has been widely studied
in the last decade and various entanglement measures have been
introduced to measure entanglement such as entanglement of
formation, entanglement of cost, relative entropy entanglement, and
negativity.

Indeed, entanglement is not the only responsible correlation for the
quantum bypassing classical regimes and there exists some quantum
correlations other than entanglement which result in the quantum
effects in quantum information processes. For instance, Bennett
\emph{et al.} showed the possibility of quantum nonlocality without
entanglement \cite{bennett99}. Also, it is shown that separable
states can be used for quantum speedup
\cite{braun99,meyer00,biham04,datta0507}. In order to quantify the
quantumness of correlations in bipartite states, it has been
proposed  several measures such as quantum discord
\cite{ollivier01}, quantum deficit
\cite{raja02,horodecki05,devetak05}, quantumness of correlations
\cite{usha08}, and quantum dissonance \cite{modi10}.

In particular, quantum discord which historically is the first
introduced measure for the nonclassicality based on the
Openheim-Horodecki paradigm, has attracted much attention in recent
years \cite{shabani09,rulli11,giorda12,terzis12,joao13}. It is based
on the difference between two quantum extensions of classically
equivalent concepts namely the mutual information. Although quantum
discord has a simple definition, its explicit evaluation is hard to
perform in practice especially for multi-qubit states and is often
only given by numerical methods. However, some analytical expression
of quantum discord for two-qubit states are presented in
Refs.~\cite{d1,d2,d3,d4,d5}.

It is known that since some separable states still have quantum
correlations, these correlations with quantum nature are more
general than entanglement. In particular, Luo \cite{luo08}
introduced a quantum-classical classification based on
measurement-induced disturbance (MID) to characterize statistical
correlations in bipartite states. In this scenario, classical states
are classified in terms of nondisturbance under quantum measurement.
However, quantum systems and thus quantum correlations are disturbed
under generic measurements and the magnitude of the disturbance can
be considered as a measure to characterize the quantumness of
states. Recently, Girolami \emph{et al.} showed that the quantum
discord due to its asymmetric definition does not properly determine
the distinction between classical-classical and classical-quantum
states and thus it is not strongly faithful. Also, by characterizing
quantum correlations in the paradigmatic instance of two-qubit
states, they observed that MID overstimates quantum correlations so
that for some classical states we obtain nonzero correlation. They
therefore proposed an ameliorated measurement-induced disturbance
(AMID) as a quantifier of quantum correlations \cite{girolami11}.

The aim of this paper is to characterize and quantify the quantum
correlation for bipartite systems which are initially prepared in
three-qubit Greenberger-Horne-Zeilinger (GHZ) \cite{ghz89}
\begin{eqnarray}
\label{ghzstate} |\mbox{GHZ}\rangle = \frac{|000\rangle +
|111\rangle}{\sqrt{2}} ,
\end{eqnarray}
and W \cite{dur00,pati}
\begin{eqnarray}
\label{wstate} |W\rangle = \frac{|100\rangle + |010\rangle+
{\sqrt{2}} |001\rangle }{2} ,
\end{eqnarray}
states under various noisy channels where the first two qubits
belong to party $a$ and the third qubit belongs to party $b$. It is
shown that this class of W states can be used for perfect
teleportation, superdense coding, and as an entanglement resource
\cite{pati}. This state belongs to the category of W states
\begin{eqnarray}
 |W_n\rangle = \frac{|100\rangle +\sqrt{n}e^{i\gamma} |010\rangle+
{\sqrt{n+1}}e^{i\delta} |001\rangle }{\sqrt{2+2n}} ,
\end{eqnarray}
where $n$ is a real number, $\delta$ and $\gamma$ are phases, and
reduces to $|W\rangle$ for $n=1$ and zero phases. Then, the initial
states are affected by noisy channels which results in decreasing of
the quantumness of states. We quantify the quantum correlations for
the initial GHZ and W states in the presence of noise and
investigate the robustness of these states under different kinds of
noise. The rest of this paper is organized as follows. In
Sec.~\ref{sec2}, we characterize quantum correlations using MID and
AMID approaches. Secs.~\ref{sec3} and \ref{sec4} are devoted to
determine quantum correlations for the GHZ and W states,
respectively. We present our conclusions in Sec.~\ref{sec5}.

\section{Classifying bipartite states using ameliorated measurement-induced disturbance}\label{sec2}
Consider a bipartite state $\rho$ for a system with two parties $a$
and $b$. Based on  measurement-induced disturbance, the quantum
correlations of $\rho$, that is denoted by ${\cal M}(\rho^{ab})$, is given by \cite{luo08}
\begin{equation}\label{qq}
{\cal M}(\rho^{ab})=I(\rho^{ab})-I(\Pi(\rho^{ab})),
\end{equation}
where $I(\rho^{ab})=S(\rho^a)+S(\rho^b)-S(\rho^{ab})$ is quantum
mutual information that quantifies the total correlation between $a$
and $b$, and
\begin{equation}\label{pi}
\Pi(\rho)=\sum_{ij}(\Pi_{a,i}\otimes\Pi_{b,j})\,\rho\,(\Pi_{a,i}\otimes\Pi_{b,j}),
\end{equation}
in which $\{\Pi_{a,i}\}$ and $\{\Pi_{b,j}\}$ are complete projective
measurements for parties $a$ and $b$, respectively. They are
obtained from the spectral decomposition of the reduced states,
namely $\rho^a=\sum_ip_{a,i}\Pi_{a,i}$ and $\rho^b=\sum_jp_{b,j}\Pi_{b,j}$.
We can rewrite Eq.~(\ref{qq}) as
\begin{equation} \label{qq2}
{\cal M}(\rho^{ab})=S(\Pi(\rho^{ab}))-S(\rho^{ab})+\sum_i(S(\rho^{x_i})-S(\Pi(\rho^{x_i}))
\end{equation}
where $x_i\in\{a,b\}$. Note that if $\Pi(\rho)=\rho$, we conclude
$\rho$ is not perturbed with respect to local measurement
$\Pi_{a,i}\otimes\Pi_{b,j}$, therefore $\rho$ is a classical state.
Otherwise, it is a quantum state and possesses quantum correlation.
For our case, since party $a$ has two qubits, it is convenient to
write Eq.~(\ref{pi}) as
$\Pi(\rho)=\sum_{ijk}(\Pi_{a,{ij}}\otimes\Pi_{b,k})\,\rho\,(\Pi_{a,{ij}}\otimes\Pi_{b,k})$.
Also, we define $\Pi_{\bf n}=\Pi_{a,{ij}}\otimes\Pi_{b,k}$ so that
the projective measurements satisfy $\Pi_{\bf n}\Pi_{{\bf
n'}}=\delta_{{\bf nn'}}\Pi_{\bf n}$ and $\sum_{\bf n}\Pi_{\bf n}=1$.

According to AMID, Eq.~(\ref{qq2}) needs to optimize over any
possible set of local projectors so that projective measurement in
this equation, which we represent by $\Omega$ instead of $\Pi$,
includes arbitrary complete projective measurements that are not
necessarily obtained from eigen-projectors. Therefore, the quantum
correlation which is denoted by ${\cal A}(\rho^{ab})$, is given by
\cite{girolami11}
\begin{equation} \label{aqq}
{\cal
A}(\rho^{ab})=\inf_{\Omega}\left[S(\Omega(\rho^{ab}))-S(\rho^{ab})+\sum_i(S(\rho^{x_i})-S(\Omega(\rho^{x_i}))\right],
\end{equation}
in which
\begin{equation}\label{omega}
\Omega(\rho)=\sum_{ij}(\Omega_{a,i}\otimes\Omega_{b,j})\,\rho\,(\Omega_{a,i}\otimes\Omega_{b,j}),
\end{equation}
$\Omega_{j,k}=U_j\Pi_{j,k} U_j^\dagger$, and $U_j=y_{j,0}I+i
\vec{\bf{y}}_j. \vec{\bf{\sigma}}_j$ is a unitary matrix obeys
$\sum_{p=0}^3y_{j,p}^2=1$, $y_{j,p}\in[-1,1]$.

The evolution of the quantum system $\rho$ in the presence of noise
is given by the master equation in the Lindblad form \cite{lind76}
\begin{equation}\label{Lindblad}
\frac{\partial \rho}{\partial t} = -\frac{i}{\hbar} [H_S, \rho] +
\sum_{i, \alpha} \left(L_{i,\alpha} \rho L_{i,\alpha}^{\dagger} -
\frac{1}{2} \left\{ L_{i,\alpha}^{\dagger} L_{i,\alpha}, \rho
\right\} \right),
\end{equation}
in which the effect of noise is presented by the Lindblad operator
$L_{i,\alpha}$ that acts on the $i$th qubit, $\alpha$ determines the
type of the noise, and $H_S$ is the Hamiltonian of the system. In
Ref.~\cite{jung08}, the authors studied analytic solutions of the
Lindblad equation for GHZ and W states under various noises for
$H_S=0$ and same axis Pauli noises by taking $L_{i,\alpha} =
\sqrt{\kappa_{i,\alpha}}\sigma^{(i)}_{\alpha}$ where
$\sigma^{(i)}_{\alpha}$ denotes Pauli noises that act on the $i$th
qubit and $\kappa$ is the decoherence rate. The time evolution of
multi-qubit GHZ states in the presence of noise is studied
analytically in Ref.~\cite{SP}.

\section{Quantumness of correlation for GHZ state}\label{sec3}
In this section, we study analytically the evolution of GHZ state
under various noisy channels and obtain corresponding quantum
correlations using the measurement-induced disturbance approach.
Also, we numerically obtain quantum correlations using AMID which
does not suffer from overestimating quantum correlations of MID
approach. The noises under investigations are the same axis Pauli
noises and the isotropic noise.

First, consider the time evolution of GHZ state in the presence of
the Pauli-X noise. For this case the solution of the Lindblad
equation reads \cite{jung08}
\begin{eqnarray}\label{ghzx}
\rho_{GHZ}^x(t)= \frac{1}{8}{\left(
\begin{matrix}
\alpha_+ & 0 & 0 & 0 & 0 & 0 & 0 & \alpha_+  \\
0 & \alpha_-  & 0& 0 & 0& 0 & \alpha_-  & 0  \\
0 & 0 & \alpha_-  & 0& 0 & \alpha_-  & 0 & 0  \\
0 & 0 & 0 & \alpha_- & \alpha_-  & 0 & 0 & 0  \\
0 & 0 & 0 & \alpha_- & \alpha_-  & 0 & 0 & 0  \\
0 & 0 & \alpha_-  & 0& 0 & \alpha_-  & 0 & 0  \\
0 & \alpha_-  & 0& 0 & 0& 0 & \alpha_- & 0  \\
\alpha_+ & 0 & 0 & 0 & 0 & 0 & 0 & \alpha_+ \end{matrix}\right),}
\end{eqnarray}
where, $\alpha_+ = 1 + 3 e ^ {-4 \kappa t}$ and $\alpha_- = 1 - e ^ {-4 \kappa t}$.

The reduced density matrices $(\rho_{GHZ}^x)^a$ and $(\rho_{GHZ}^x)^b$ are found by
tracing out the third qubit and the first two qubits, respectively,
\begin{eqnarray}
(\rho_{GHZ}^x)^a= \frac{1}{8}{\left(
\begin{matrix}
\alpha_+ + \alpha_- & 0 & 0 & 0  \\
0 & 2\alpha_- & 0 & 0  \\
0 & 0 &2 \alpha_- & 0  \\
0 & 0 & 0 & \alpha_+ + \alpha_- \end{matrix}\right)},\hspace{1cm}
(\rho_{GHZ}^x)^b= \frac{\alpha_++3\alpha_-}{8}I.
\end{eqnarray}
Thus, the projective measurements are given by $\Pi_{a,{ij}}=|
ij\rangle\langle ij|$, $\Pi_{b,k}=| k\rangle\langle k|$, and
$\Pi(\rho_{GHZ}^x)$ reads
\begin{eqnarray}
\Pi(\rho_{GHZ}^x)=
\frac{1}{8}{\left(
\begin{matrix}
\alpha_+ & 0 & 0 & 0 & 0 & 0 & 0 & 0  \\
0 & \alpha_-  & 0& 0 & 0& 0 & 0 & 0  \\
0 & 0 & \alpha_-  & 0& 0 & 0  & 0 & 0  \\
0 & 0 & 0 & \alpha_- & 0 & 0 & 0 & 0  \\
0 & 0 & 0 & 0 & \alpha_-  & 0 & 0 & 0  \\
0 & 0 & 0 & 0& 0 & \alpha_-  & 0 & 0  \\
0 & 0 & 0& 0 & 0& 0 & \alpha_- & 0  \\
0 & 0 & 0 & 0 & 0 & 0 & 0 & \alpha_+ \end{matrix}\right)}.
\end{eqnarray}
Since $\left[\Pi(\rho_{GHZ}^x)\right]^a=\rho^a$ and
$\left[\Pi(\rho_{GHZ}^x)\right]^b=\rho^b$ the third term in
Eq.~(\ref{qq2}) vanishes  and we only need to evaluate
$S(\rho_{GHZ}^x)$ and $S(\Pi(\rho_{GHZ}^x))$, namely
\begin{eqnarray} \label{entropy1}
S(\rho_{GHZ}^x)=2-\frac{\alpha_+}{4}\log_2(\alpha_+)-\frac{3\alpha_-}{4}\log_2(\alpha_-),
\end{eqnarray}
and
\begin{eqnarray} \label{entropy2}
S(\Pi(\rho_{GHZ}^x))=3-\frac{\alpha_+}{4}\log_2(\alpha_+)-\frac{3\alpha_-}{4}\log_2(\alpha_-).
\end{eqnarray}
Therefore, the quantum correlation of $\rho_{GHZ}^x$ is given by
\begin{eqnarray}
{\cal M}(\rho_{GHZ}^x)=1.
\end{eqnarray}

\begin{figure}
\begin{center}
\includegraphics[width=10cm]{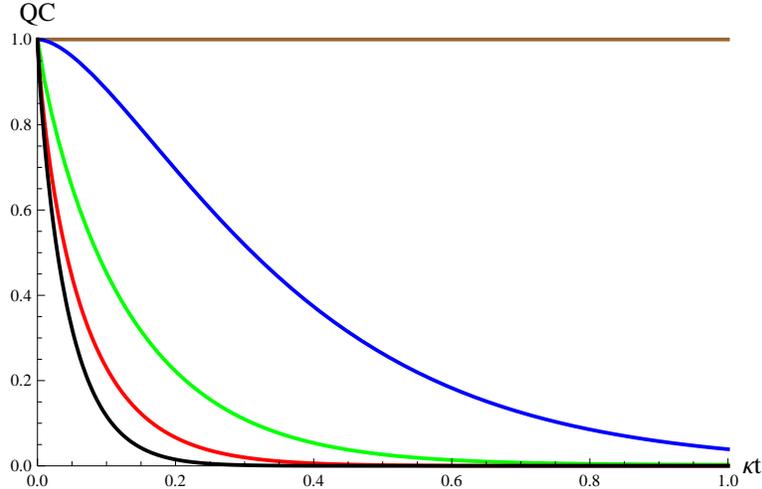}
\caption{\label{fig1} Quantum correlation evaluated by MID and AMID
for the three-qubit system with the initial GHZ state as a function
of $\kappa t$ transmitted through various noisy channels: Pauli-X by
MID (brown line), Pauli-X by AMID (green line), Pauli-Y by MID and
AMID (blue line), Pauli-Z by MID and AMID (red line), and isotropic
by MID and AMID (black line).}
\end{center}
\end{figure}

Now, in order to obtain quantum correlations by AMID, we need to
evaluate Eq.~(\ref{aqq}) for the density matrix $\rho_{GHZ}^x(t)$
(\ref{ghzx}). For this purpose, first we construct the unitary
matrices $U_j$ by choosing $y_{j,0}=\cos\psi_j$, $y_{j,1}
=\sin\psi_j \cos\theta_j$, $y_{j,2}
=\sin\psi_j\sin\theta_j\sin\phi_j$, $y_{j,3}
=\sin\psi_j\sin\theta_j\cos\phi_j$ that satisfy
$\sum_{p=0}^3y_{j,p}^2=1$. Then, we find $\Omega(\rho)$ and obtain
the corresponding von-Neumann entropies in Eq.~(\ref{aqq}). Thus,
the quantum correlation is found as a function of nine parameters
and time, i.e., ${\cal
A}(\theta_1,\phi_1,\psi_1,\theta_2,\phi_2,\psi_2,\theta_3,\phi_3,\psi_3,\kappa
t)$. Now, the optimization program over these nine parameters gives
rise to AMID. For this case, we have ${\cal A}={\cal
A}(1.3,4.43,2.31,1.3,4.43,2.31,1.3,4.43,2.31,\kappa t)$ which is
depicted in Fig.~\ref{fig1} (green line). As the figure
shows, although ${\cal M}=1$ for all times in presence of a bit-flip
noise, the AMID represents dissipative behavior for the quantum
correlation.

For the Pauli-Y noise the density matrix reads \cite{jung08}
\begin{eqnarray}
\rho_{GHZ}^y(t)= \frac{1}{8}{\left(
\begin{matrix}
\alpha_+ & 0 & 0 & 0 & 0 & 0 & 0 & \beta_1  \\
0 & \alpha_-  & 0& 0 & 0& 0 & -\beta_2  & 0  \\
0 & 0 & \alpha_-  & 0& 0 & -\beta_2  & 0 & 0  \\
0 & 0 & 0 & \alpha_- & -\beta_2  & 0 & 0 & 0  \\
0 & 0 & 0 & -\beta_2 & \alpha_-  & 0 & 0 & 0  \\
0 & 0 & -\beta_2  & 0& 0 & \alpha_-  & 0 & 0  \\
0 & -\beta_2  & 0& 0 & 0& 0 & \alpha_- & 0  \\
\beta_1 & 0 & 0 & 0 & 0 & 0 & 0 & \alpha_+ \end{matrix}\right)},
\end{eqnarray}
where $\beta_1 = 3 e ^ {-2 \kappa t} + e ^ {-6 \kappa t}$ and
$\beta_2 = e ^ {-2 \kappa t} - e ^ {-6 \kappa t}$.

It is straightforward to check that the reduced density matrices,
the projective measurements, and $\Pi(\rho_{GHZ}^y)$ are similar to
the previous case. So, to obtain ${\cal M}(\rho_{GHZ}^y)$ we only require
to evaluate $S(\rho_{GHZ}^y)$ as
\begin{eqnarray}\label{entropy3}
S(\rho_{GHZ}^y)= 3 &-&\frac{\alpha_+ - \beta_1}{8}\log_2(\alpha_+ -
\beta_1)-\frac{\alpha_+ + \beta_1}{8}\log_2(\alpha_+ +
\beta_1)\nonumber \\ &-&3\frac{\alpha_- - \beta_2}{8}\log_2(\alpha_-
- \beta_2)-3\frac{\alpha_- + \beta_2}{8}\log_2(\alpha_- + \beta_2).
\end{eqnarray}
Now, the quantum correlation is
\begin{eqnarray}\label{qc:y}
{\cal M}(\rho_{GHZ}^y)&=& \frac{\alpha_+ - \beta_1}{8}\log_2(\alpha_+ -
\beta_1)+\frac{\alpha_+ + \beta_1}{8}\log_2(\alpha_+ +
\beta_1)\nonumber \\
&&+3\frac{\alpha_- - \beta_2}{8}\log_2(\alpha_- -
\beta_2)+3\frac{\alpha_- + \beta_2}{8}\log_2(\alpha_- + \beta_2)
\nonumber \\
&&-\frac{\alpha_+}{4}\log_2(\alpha_+)-3\frac{\alpha_-}{4}\log_2(\alpha_-).
\end{eqnarray}
The optimization procedure based on AMID shows that for this case
the quantum correlation calculated by both measures coincide, i.e.,
${\cal A}={\cal M}$.

For the Pauli-Z noise, GHZ state under the noisy channel is
described by \cite{jung08}
\begin{eqnarray}
\rho_{GHZ}^z(t) = \frac{1}{2} \left( |000\rangle \langle 000 | +
|111\rangle\langle 111| \right) + \frac{1}{2}e ^ {-6 \kappa
t}\left(|000\rangle
 \langle 111| + |111\rangle \langle000 | \right),
\end{eqnarray}
and the reduced density matrices are
\begin{eqnarray}\label{roaz}
(\rho_{GHZ}^z)^a= \frac{1}{2} \left( |00\rangle \langle 00 | +
|11\rangle\langle 11| \right), \hspace{1cm}\label{robz} (\rho_{GHZ}^z)^b=
\frac{1}{2}I.
\end{eqnarray}
The projective measurements are similar to the previous cases and
\begin{eqnarray}
\Pi(\rho_{GHZ}^z)= \frac{1}{2}\left( |000\rangle \langle 000 | +
|111\rangle\langle 111| \right),
\end{eqnarray}
which results in the unity of the corresponding von-Neumann entropy,
i.e., $S(\Pi(\rho_{GHZ}^z))=1$. Tracing out the third qubit and the
first two qubits leads to Eqs.~($\ref{roaz}$). So, we have
$S((\rho_{GHZ}^z)^{x_i})-S(\Pi(\rho_{GHZ}^z)^{x_i})=0$ with $x_i\in\{a,b\}$ and the
third term in Eq.~(\ref{qq2}) vanishes.

Now, in order to evaluate the quantum correlation, we find the
von-Neumann entropy of $\rho_{GHZ}^z$
\begin{eqnarray}
S(\rho_{GHZ}^z)=1-\frac{1- e^{-6 \kappa t}}{2}\log_2(1- e^{-6 \kappa t})-\frac{1 + e^{-6 \kappa t}}{2}\log_2(1 + e^{-6 \kappa t}),
\end{eqnarray}
which results in
\begin{eqnarray}
{\cal M}(\rho_{GHZ}^z)=\frac{1- e^{-6 \kappa t}}{2}\log_2(1- e^{-6 \kappa t})+\frac{1 + e^{-6 \kappa t}}{2}\log_2(1 + e^{-6 \kappa t}).
\end{eqnarray}
Applying unitary matrices on the projective bases to get the local
projective measurements and computing the von-Neumann entropies of
Eq.~(\ref{aqq}) result in ${\cal A}$. It is found that the obtained
optimized quantum correlation agrees with ${\cal M}(\rho_{GHZ}^z)$.

For the last case in this section, we investigate the GHZ state
which is affected by the isotropic noise. Its density matrix is
given by \cite{jung08}
\begin{eqnarray}
\rho_{GHZ}^d(t)= \frac{1}{8}{\left(
\begin{matrix}
\tilde{\alpha}_+ & 0 & 0 & 0 & 0 & 0 & 0 & \gamma  \\
0 & \tilde{\alpha}_-  & 0& 0 & 0& 0 & 0 & 0  \\
0 & 0 & \tilde{\alpha}_-  & 0& 0 & 0  & 0 & 0  \\
0 & 0 & 0 & \tilde{\alpha}_- & 0 & 0 & 0 & 0  \\
0 & 0 & 0 & 0 & \tilde{\alpha}_-  & 0 & 0 & 0  \\
0 & 0 & 0 & 0& 0 & \tilde{\alpha}_-  & 0 & 0  \\
0 & 0 & 0& 0 & 0& 0 & \tilde{\alpha}_- & 0  \\
\gamma & 0 & 0 & 0 & 0 & 0 & 0 & \tilde{\alpha}_+
\end{matrix}\right),}
\end{eqnarray}
where  $\tilde{\alpha}_+ =1 + 3e ^ {-8 \kappa t}$, $\tilde{\alpha}_-
= 1 - e ^ {-8 \kappa t}$ and $\gamma = 4 e ^ {-12 \kappa t}$. The
reduced density matrices for the subsystems $a$ and $b$ are
\begin{eqnarray}
(\rho_{GHZ}^d)^a=
\frac{1}{8}{\left(
\begin{matrix}
\tilde{\alpha}_++\tilde{\alpha}_- & 0 & 0 & 0  \\
0 & \tilde{\alpha}_- & 0 & 0  \\
0 & 0 & \tilde{\alpha}_- & 0  \\
0 & 0 & 0 & \tilde{\alpha}_++\tilde{\alpha}_- \end{matrix}\right),}
\hspace{1cm}\label{robz1} (\rho_{GHZ}^d)^b=
\frac{\tilde{\alpha}_++3\tilde{\alpha}_-}{8}I.
\end{eqnarray}
Thus, the projective measurements are again given by $\Pi_{a,{ij}}=|
ij\rangle\langle ij|$, $\Pi_{b,k}=| k\rangle\langle k|$, and we find
\begin{eqnarray}
\Pi(\rho_{GHZ}^d)=
\frac{1}{8}{\left(
\begin{matrix}
\tilde{\alpha}_+ & 0 & 0 & 0 & 0 & 0 & 0 & 0  \\
0 & \tilde{\alpha}_-  & 0& 0 & 0& 0 & 0 & 0  \\
0 & 0 & \tilde{\alpha}_-  & 0& 0 & 0  & 0 & 0  \\
0 & 0 & 0 & \tilde{\alpha}_- & 0 & 0 & 0 & 0  \\
0 & 0 & 0 & 0 & \tilde{\alpha}_-  & 0 & 0 & 0  \\
0 & 0 & 0 & 0& 0 & \tilde{\alpha}_-  & 0 & 0  \\
0 & 0 & 0& 0 & 0& 0 & \tilde{\alpha}_- & 0  \\
0 & 0 & 0 & 0 & 0 & 0 & 0 & \tilde{\alpha}_+ \end{matrix}\right),}
\end{eqnarray}
which results in $\left[\Pi(\rho_{GHZ}^d)\right]^a=(\rho_{GHZ}^d)^a$ and
$\left[\Pi(\rho_{GHZ}^d)\right]^b=(\rho_{GHZ}^d)^b$. Therefore,
$S((\rho_{GHZ}^d)^{x_i})-S(\Pi(\rho_{GHZ}^d)^{x_i})=0$ and Eq.~($\ref{qq2}$)
reduces to
\begin{equation}\label{qq3}
{\cal M}(\rho_{GHZ}^d)=S(\Pi(\rho_{GHZ}^d))-S(\rho_{GHZ}^d).
\end{equation}
The von-Neumann entropies are given by
\begin{eqnarray}\label{en5}
S(\rho_{GHZ}^d)=3-3\frac{\tilde{\alpha}_-}{4}\log_2(\tilde{\alpha}_-)-\frac{
\tilde{\alpha}_+-\gamma}{8}\log_2(\tilde{\alpha}_+-\gamma)-\frac{
\tilde{\alpha}_++\gamma}{8}\log_2(\tilde{\alpha}_++\gamma),
\end{eqnarray}
and
\begin{eqnarray}\label{en6}
S(\Pi(\rho_{GHZ}^d))=3-\frac{\tilde{\alpha}_+}{4}\log_2(\tilde{\alpha}_+)-\frac{3\tilde{\alpha}_-}{4}\log_2(\tilde{\alpha}_-).
\end{eqnarray}
So, the quantum correlation reads
\begin{eqnarray}\label{qq4}
{\cal M}(\rho_{GHZ}^d)=\frac{
\tilde{\alpha}_++\gamma}{8}\log_2(\tilde{\alpha}_++\gamma)+\frac{
\tilde{\alpha}_+-\gamma}{8}\log_2(\tilde{\alpha}_+-\gamma)-\frac{\tilde{\alpha}_+}{4}\log_2(\alpha_+).
\end{eqnarray}
Similar to the previous case, the quantum correlation obtained by
AMID coincides with one obtained by MID for all times.

In Fig.~\ref{fig1} we have depicted the quantum correlation obtained
by MID and AMID  for GHZ state in the presence of various noisy
channels. As the figure shows, quantum correlations for all noises,
except Pauli-X noise, coincide for both measures. MID overestimates
quantum correlation for Pauli-X channel with respect to AMID
measure. Note that, for the GHZ state ${\cal M}(\rho)$ for all
noises that are studied in this contribution agrees with the
corresponding asymmetric quantum discord \cite{mah12}.

\section{Quantumness of correlation for W state}\label{sec4}
In this section, we determine quantum correlation using MID and AMID for a bipartite
state which is initially prepared in the form of W state under
various noise channels. The first two qubits belong to party $a$ and
the third qubit belongs to party $b$.

\begin{figure}
\begin{center}
\includegraphics[width=10cm]{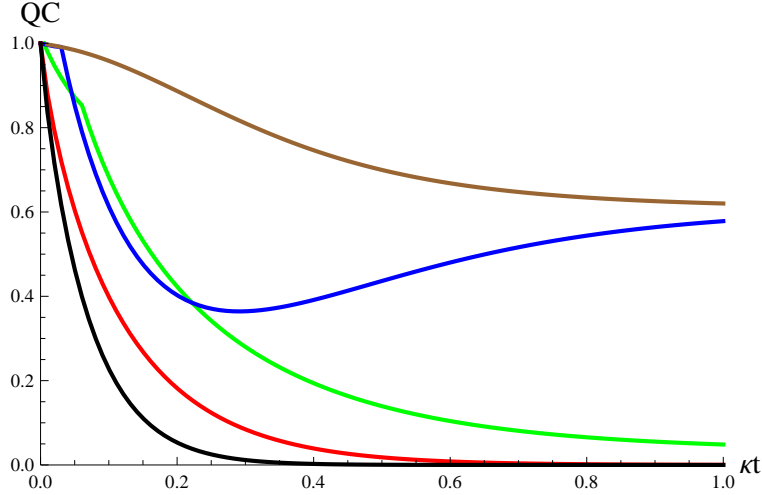}
\caption{\label{fig2}   Quantum correlation evaluated by MID and
AMID for the three-qubit system with the initial W state as a
function of $\kappa t$ transmitted through various noisy channels:
Pauli-X and Pauli-Y by MID (brown line), Pauli-X by AMID (green
line), Pauli-Y by AMID (blue line), Pauli-Z by MID and AMID (red
line), and isotropic by MID and AMID (black line).}
\end{center}
\end{figure}

In the presence of the Pauli-X noise, the time evolution of the density matrix of W state is given by \cite{jung08}
\begin{eqnarray}\label{wx}
\rho_{W}^x(t)=
\frac{1}{16}{\left(
\begin{matrix}
2\alpha_2 & 0 & 0 & \sqrt{2}\alpha_2 & 0 &  \sqrt{2}\alpha_2 & \alpha_2 & 0 \\
0 & 2\alpha_1 & \sqrt{2}\alpha_1 & 0 & \sqrt{2}\alpha_1 & 0 & 0 & \alpha_3  \\
0 & \sqrt{2}\alpha_1  & 2\beta_+ & 0& \alpha_1 & 0  & 0 & \sqrt{2}\alpha_3  \\
\sqrt{2}\alpha_2 & 0 & 0 & 2\beta_- & 0 & \alpha_4 & \sqrt{2}\alpha_4 & 0  \\
0 & \sqrt{2}\alpha_1 & \alpha_1 & 0 & 2\beta_+ & 0 & 0 & \sqrt{2}\alpha_3   \\
\sqrt{2}\alpha_2  & 0 & 0 & \alpha_4 & 0 & 2\beta_- & \sqrt{2}\alpha_4 & 0  \\
\alpha_2 & 0 & 0 & \sqrt{2}\alpha_4 & 0 & \sqrt{2}\alpha_4 & 2\alpha_4 & 0  \\
0 & \alpha_3 & \sqrt{2}\alpha_3 & 0 & \sqrt{2}\alpha_3 & 0 & 0 & 2\alpha_3 \end{matrix}\right),}
\end{eqnarray}
where
\begin{eqnarray}\left\{
\begin{array}{l}
\alpha_1=1 + e^ {-2\kappa t}+ e^{-4\kappa t}+ e^{-6\kappa t},\\
\alpha_2=1 + e^ {-2\kappa t}- e^{-4\kappa t}- e^{-6\kappa t},\\
\alpha_3=1 - e^ {-2\kappa t}- e^{-4\kappa t}+ e^{-6\kappa t},\\
\alpha_4=1 - e^ {-2\kappa t}+ e^{-4\kappa t}- e^{-6\kappa t},\\
\beta_\pm =1 \pm e^{-6\kappa t}.\\
\end{array}\right.
\end{eqnarray}
The projective measurements are found using the reduced density matrices
\begin{eqnarray}\label{rhoawx}
(\rho_{W}^x)^a=
\frac{1}{16}{\left(
\begin{matrix}
2(\alpha_1+\alpha_2) & 0 & 0 & \alpha_2 + \alpha_3  \\
0 & 2(\beta_+ + \beta_-) & \alpha_1+\alpha_4 & 0  \\
0 & \alpha_1+\alpha_4 & 2(\beta_+ + \beta_-) & 0  \\
\alpha_2 + \alpha_3  & 0 & 0 & 2(\alpha_3 + \alpha_4) \end{matrix}\right),}\hspace{1cm}
(\rho_{W}^x)^b=
\frac{I}{2},
\end{eqnarray}
which results in
\begin{eqnarray}\left\{
\begin{array}{l}
\Pi_{a,{00}}=\frac{1}{2}(|01\rangle + \langle 10|)( \langle 01| + \langle 10|),\\
\Pi_{a,{11}}=\frac{1}{2}(|01\rangle - \langle 10|)( \langle 01| - \langle 10|),\\
\Pi_{a,{01}}=\frac{1}{2(1+e^{4\kappa t})}\left((1-e^{2\kappa t})^2|00\rangle\langle 00| + (1-e^{4\kappa t}) |11\rangle\langle 11|\right),\\
\Pi_{a,{10}}=\frac{1}{2(1+e^{4\kappa t})}\left((1+e^{2\kappa t})^2|00\rangle\langle 00| - (1-e^{4\kappa t}) |11\rangle\langle 11|\right),\\
\end{array}\right.
\end{eqnarray}
and $\Pi_{b,i}=|i\rangle\langle i|$. So we have
\begin{equation}
\Pi(\rho_W^x)={\left(
\begin{matrix}
\gamma_1 & 0 & 0 & 0 & 0 & 0 & \eta_1 & 0 \\
0 & \gamma_2 & 0 & 0 & 0 & 0 & 0 & \eta_2  \\
0 & 0  & 2\beta_+ & 0& \alpha_1 & 0 & 0 & 0  \\
0 & 0 & 0 & 2\beta_- & 0 & \alpha_4 & 0 & 0  \\
0 & 0 & \alpha_1 & 0 & 2\beta_+ & 0 & 0 & 0  \\
0 & 0 & 0 & \alpha_4  & 0 & 2\beta_- & 0 & 0  \\
\eta_1 & 0 & 0 & 0 & 0 & 0 & \gamma_3 & 0  \\
0 & \eta_2 & 0 & 0 & 0 & 0 & 0 & \gamma_4
\end{matrix} \right),}
\end{equation}
where
\begin{eqnarray}\left\{
\begin{array}{l}
\gamma_1=\frac{2 e^{\kappa t}\cosh(\kappa t)^2\sinh(\kappa t)(2-\cosh(2\kappa t)+\cosh(4\kappa t)+\sinh(2\kappa t))}{(1+e^{4\kappa t})^2},\\
\gamma_2=\frac{1+e^{-6\kappa t}+\frac{8}{(1+e^{4\kappa t})^2}+2e^{-3\kappa t}\sinh(\kappa t)}{8},\\
\gamma_3=\frac{1-e^{-6\kappa t}+\frac{8}{(1+e^{4\kappa t})^2}-2e^{-3\kappa t}\cosh(\kappa t)}{8},\\
\gamma_4=\frac{2 e^{\kappa t}\cosh(\kappa t)\sinh(\kappa t)^2(2+\cosh(2\kappa t)+\cosh(4\kappa t)-\sinh(2\kappa t))}{(1+e^{4\kappa t})^2},\\
\eta_1=\frac{ e^{\kappa t}\sinh(\kappa t)\sinh(2\kappa t)(2+2\sinh(2\kappa t)+\sinh(4\kappa t))}{2(1+e^{4\kappa t})^2},\\
\eta_2=\frac{ e^{\kappa t}\cosh(\kappa t)^2\sinh(\kappa t)(2-2\sinh(2\kappa t)+\sinh(4\kappa t))}{(1+e^{4\kappa t})^2}.
\end{array}\right.
\end{eqnarray}
Since $\left[\Pi(\rho_{W}^x)\right]^a=(\rho_{W}^x)^a$ and
$\left[\Pi(\rho_{W}^x)\right]^b=(\rho_{W}^x)^b$, the third term of Eq.~($\ref{qq2}$) vanishes and we obtain
\begin{equation}\label{qq5}
{\cal M}(\rho_W^x)=S(\Pi(\rho_{W}^x))-S(\rho_{W}^x).
\end{equation}
Now, the quantum correlation can be evaluated numerically which is depicted in
Fig.~\ref{fig2}.

The numerical optimization program for the nine parameters that is
inherent in AMID approach gives rise to the following nonclassical
correlation
\begin{equation}
{\cal A}=\left\{ {\hspace{-0.7cm}{{\cal
A}(2.23,0,1.1,1.1,0,1.1,1.1,0,1.1,\kappa t),} \atop {{\cal
A}(2.2,2.3,2.2,2.2,2.3,2.2,2.2,2.3,2.2,\kappa t),}} \hskip 1cm
{{0<\kappa t<0.06,} \atop {\kappa t>0.06,}} \right.
\end{equation}
which is depicted in Fig.~\ref{fig2} as a green line.

For the Pauli-Y noise the density matrix reads \cite{jung08}
\begin{eqnarray}
\rho_{W}^y(t)=
\frac{1}{16}{\left(
\begin{matrix}
2\alpha_2 & 0 & 0 & -\sqrt{2}\alpha_2 & 0 &  -\sqrt{2}\alpha_2 & -\alpha_2 & 0 \\
0 & 2\alpha_1 & \sqrt{2}\alpha_1 & 0 & \sqrt{2}\alpha_1 & 0 & 0 & -\alpha_3  \\
0 & \sqrt{2}\alpha_1  & 2\beta_+ & 0& \alpha_1 & 0  & 0 & -\sqrt{2}\alpha_3  \\
-\sqrt{2}\alpha_2 & 0 & 0 & 2\beta_- & 0 & \alpha_4 & \sqrt{2}\alpha_4 & 0  \\
0 & \sqrt{2}\alpha_1 & \alpha_1 & 0 & 2\beta_+ & 0 & 0 & -\sqrt{2}\alpha_3   \\
-\sqrt{2}\alpha_2  & 0 & 0 & \alpha_4 & 0 & 2\beta_- & \sqrt{2}\alpha_4 & 0  \\
-\alpha_2 & 0 & 0 & \sqrt{2}\alpha_4 & 0 & \sqrt{2}\alpha_4 & 2\alpha_4 & 0  \\
0 & -\alpha_3 & -\sqrt{2}\alpha_3 & 0 & -\sqrt{2}\alpha_3 & 0 & 0 & 2\alpha_3 \end{matrix}\right).}
\end{eqnarray}
For this case the results are identical with the previous case.
Therefore, the time evolution of ${\cal M}$ for the initial W state
under Pauli-Y noise coincides with ${\cal M}(\rho_W^x)$ (see
Fig.~\ref{fig2}). Also, the evolution of quantum correlation
computed by  AMID results in
\begin{equation}
{\cal A}=\left\{
{{{\cal A}(1.57,1.57,1.57,1.57,1.57,1.57,1.57,1.57,1.57,\kappa t)} \atop {{\cal A}(1.57,2.22,1.57,1.57,2.22,1.57,1.57,2.22,1.57,\kappa t)}} \hskip 1cm
{{0<\kappa t<0.03} \atop {\kappa t>0.03,}}
\right.
\end{equation}
which is shown in Fig.~\ref{fig2} as a blue line.

To this end, consider the effects of the Pauli-Z noise on W state
\cite{jung08}
\begin{eqnarray}\label{wz}
\rho_{W}^z(t)=
\frac{1}{4}{\left(
\begin{matrix}
0 & 0 & 0 & 0 & 0 & 0 & 0 &0 \\
0 & 2 & \sqrt{2}e^{-4\kappa t}& 0 & \sqrt{2}e^{-4\kappa t} & 0 & 0 & 0  \\
0 & \sqrt{2}e^{-4\kappa t} & 1 & 0& e^{-4\kappa t} & 0  & 0 & 0  \\
0 & 0 & 0 & 0 & 0 & 0 & 0 & 0  \\
0 & \sqrt{2}e^{-4\kappa t} & e^{-4\kappa t} & 0 & 1 & 0 & 0 & 0  \\
0 & 0 & 0 & 0 & 0 & 0 & 0 & 0  \\
0 & 0 & 0 & 0 & 0 & 0 & 0 & 0  \\
0 & 0 & 0 & 0 & 0 & 0 & 0 & 0 \end{matrix}\right).}
\end{eqnarray}
So, the reduced density matrices for the subsystems are
\begin{eqnarray}\label{rhoawz}
(\rho_{W}^z)^a=
\frac{1}{4}{\left(
\begin{matrix}
2 & 0 & 0 & 0  \\
0 & 1 & e^{-4\kappa t} & 0  \\
0 & e^{-4\kappa t} & 1 & 0  \\
0 & 0 & 0 & 0 \end{matrix}\right),}\hspace{1cm}
(\rho_{W}^z)^b=
\frac{I}{2},
\end{eqnarray}
which result in
\begin{eqnarray}\left\{
\begin{array}{l}
\Pi_{a,{00}}=|00\rangle \langle 00|,\\
\Pi_{a,{11}}=|11\rangle \langle 11|,\\
\Pi_{a,{01}}=\frac{1}{2}(|01\rangle + \langle 10|)( \langle 01| + \langle 10|),\\
\Pi_{a,{10}}=\frac{1}{2}(|01\rangle - \langle 10|)( \langle 01| - \langle 10|),\\
\end{array}\right.
\end{eqnarray}
and $\Pi_{b,i}=|i\rangle \langle i|$.
Using the projective measurements we find
\begin{eqnarray}
\Pi(\rho_{W}^z)=
\frac{1}{4}{\left(
\begin{matrix}
0 & 0 & 0 & 0 & 0 & 0 & 0 &0 \\
0 & 2 & 0 & 0 & 0 & 0 & 0 & 0  \\
0 & 0 & 1 & 0& e^{-4\kappa t} & 0  & 0 & 0  \\
0 & 0 & 0 & 0 & 0 & 0 & 0 & 0  \\
0 & 0 & e^{-4\kappa t} & 0 & 1 & 0 & 0 & 0  \\
0 & 0 & 0 & 0 & 0 & 0 & 0 & 0  \\
0 & 0 & 0 & 0 & 0 & 0 & 0 & 0  \\
0 & 0 & 0 & 0 & 0 & 0 & 0 & 0 \end{matrix}\right),}
\end{eqnarray}
and $\left[\Pi(\rho_{W}^z)\right]^a=(\rho_{W}^z)^a$ and
$\left[\Pi(\rho_{W}^z)\right]^b=(\rho_{W}^z)^b$.  Therefore, we only need to obtain the following von-Neumann entropies
\begin{eqnarray}
S(\rho_W^z)&=&\frac{1}{4}(11+e^{-4\kappa
t})-\frac{1}{4}(1-e^{-4\kappa t})\log_2(1-e^{-4\kappa t})\nonumber \\
&-&\frac{1}{8}[(3+e^{-4\kappa t}-\sqrt{1-2e^{-4\kappa
t}+17e^{-8\kappa t}})\log_2(3+e^{-4\kappa t}-\sqrt{1-2e^{-4\kappa
t}+17e^{-8\kappa t}})\nonumber \\  &+&(3+e^{-4\kappa
t}+\sqrt{1-2e^{-4\kappa t}+17e^{-8\kappa t}})\log_2(3+e^{-4\kappa
t}+\sqrt{1-2e^{-4\kappa t}+17e^{-8\kappa t}})],\hspace{0.7cm}
\end{eqnarray}
and
\begin{eqnarray}
S(\Pi(\rho_W^z))=\frac{3}{2}-\frac{1}{4}(1-e^{-4\kappa t})\log_2(1-e^{-4\kappa t})-\frac{1}{4}(1+e^{-4\kappa t})\log_2(1+e^{-4\kappa t}).
\end{eqnarray}
Now, the quantum correlation can be found analytically
\begin{eqnarray}
\hspace{-0.2cm}{\cal M}(\rho_W^z)&=&-\frac{1}{4}(5+e^{-4\kappa t})-\frac{1}{4}(1+e^{-4\kappa t})\log_2(1+e^{-4\kappa t})\nonumber \\
&+&\frac{1}{8}[(3+e^{-4\kappa t}-\sqrt{1-2e^{-4\kappa t}+17e^{-8\kappa t}})\log_2(3+e^{-4\kappa t}-\sqrt{1-2e^{-4\kappa t}+17e^{-8\kappa t}})\nonumber \\
&+&(3+e^{-4\kappa t}+\sqrt{1-2e^{-4\kappa t}+17e^{-8\kappa t}})\log_2(3+e^{-4\kappa t}+\sqrt{1-2e^{-4\kappa t}+17e^{-8\kappa t}})].\hspace{0.7cm}
\end{eqnarray}
Performing the optimization procedure of AMID gives us the same
result, namely, ${\cal
A}(\theta_1,\phi_1,0,\theta_2,\phi_2,0,\theta_3,\phi_3,0)={\cal
M}(\rho_W^z)$. In other words, the infimum value of  ${\cal A}$
happens for $\psi_i=0$ $(i=1,2,3)$ and arbitrary values of $\phi_i$
and $\theta_i$.

For the last case, consider the isotropic noise. The corresponding
density matrix reads \cite{jung08}
\begin{eqnarray}
\rho_{W}^d(t)=
\frac{1}{8}{\left(
\begin{matrix}
\tilde{\alpha}_2 & 0 & 0 & 0 & 0 & 0 & 0 & 0 \\
0 & \tilde{\alpha}_1 & \sqrt{2}\tilde{\gamma}_+ & 0 & \sqrt{2}\tilde{\gamma}_+ & 0 & 0 & 0  \\
0 & \sqrt{2}\tilde{\gamma}_+ & \tilde{\beta}_+ & 0& \tilde{\gamma}_+ & 0  & 0 & 0  \\
0 & 0 & 0 & \tilde{\beta}_- & 0 & \tilde{\gamma}_- & \sqrt{2}\tilde{\gamma}_- & 0  \\
0 & \sqrt{2}\tilde{\gamma}_+ & \tilde{\gamma}_+ & 0 & \tilde{\beta}_+ & 0 & 0 & 0  \\
0 & 0 & 0 & \tilde{\gamma}_- & 0 & \tilde{\beta}_- & \sqrt{2}\tilde{\gamma}_- & 0  \\
0 & 0 & 0 & \sqrt{2}\tilde{\gamma}_- & 0 & \sqrt{2}\tilde{\gamma}_- & \tilde{\alpha}_4 & 0  \\
0 & 0 & 0 & 0 & 0 & 0 & 0 &\tilde{\alpha}_3 \end{matrix}\right),}
\end{eqnarray}
where
\begin{eqnarray}\left\{
\begin{array}{l}
\tilde\alpha_1=1 + e^ {-4\kappa t}+ e^{-8\kappa t}+ e^{-12\kappa t},\\
\tilde\alpha_2=1 + e^ {-4\kappa t}- e^{-8\kappa t}- e^{-12\kappa t},\\
\tilde\alpha_3=1 - e^ {-4\kappa t}- e^{-8\kappa t}+ e^{-12\kappa t},\\
\tilde\alpha_4=1 - e^ {-4\kappa t}+ e^{-8\kappa t}- e^{-12\kappa t},\\
\tilde\gamma_\pm =1 \pm e^{-6\kappa t}.\\
\end{array}\right.
\end{eqnarray}
The reduced density matrices are given by
\begin{eqnarray}\label{awd}
(\rho_{W}^d)^a=
\frac{1}{8}{\left(
\begin{matrix}
\tilde{\alpha_1}+\tilde{\alpha_2} & 0 & 0 & 0  \\
0 & \tilde{\beta_-}+\tilde{\beta_+} & \tilde{\gamma_-}+\tilde{\gamma_+} & 0  \\
0 & \tilde{\gamma_-}+\tilde{\gamma_+} & \tilde{\beta_-}+\tilde{\beta_+} & 0  \\
0 & 0 & 0 & \tilde{\alpha_3}+\tilde{\alpha_4} \end{matrix}\right),}\hspace{.5cm}
(\rho_{W}^d)^b=
\frac{1}{8}{\left(
\begin{matrix}
\tilde{\beta_+}+\tilde{\alpha_2}+\tilde{\alpha_4} & 0  \\
0 & \tilde{\beta_-}+\tilde{\alpha_1}+\tilde{\alpha_3} \end{matrix}\right),}\hspace{.5cm}
\end{eqnarray}
and
\begin{eqnarray}
\Pi(\rho_{W}^d)=
\frac{1}{8}{\left(
\begin{matrix}
\tilde{\alpha_2} & 0 & 0 & 0 & 0 & 0 & 0 & 0 \\
0 & \tilde{\alpha_1} & 0 & 0 & 0 & 0 & 0 & 0  \\
0 & 0 & \tilde{\beta_+} & 0 & \tilde{\gamma_+} & 0  & 0 & 0  \\
0 & 0 & 0 & \tilde{\beta_-} & 0 & \tilde{\gamma_-} & 0 & 0  \\
0 & 0 & \tilde{\gamma_+} & 0 & \tilde{\beta_+} & 0 & 0 & 0  \\
0 & 0 & 0 & \tilde{\gamma_-} & 0 & \tilde{\beta_-} & 0 & 0  \\
0 & 0 & 0 & 0 & 0 & 0 & \tilde{\alpha_4} & 0  \\
0 & 0 & 0 & 0 & 0 & 0 & 0 & \tilde{\alpha_3} \end{matrix}\right)}.
\end{eqnarray}
Tracing out the first two qubits and the third qubit leads to
$\left[\Pi(\rho_{W}^d)\right]^a=(\rho_{W}^d)^a$ and
$\left[\Pi(\rho_{W}^d)\right]^b=(\rho_{W}^d)^b$ which results in
$S((\rho_W^d)^a)=S(\Pi(\rho_W^d)^a)$ and
$S((\rho_W^d)^b)=S(\Pi(\rho_W^d)^b)$. Thus, using the von-Neumann
entropies
\begin{eqnarray}\label{en7}
S(\rho_W^d)&=&\frac{1}{2}(7+e^{-8\kappa t})-\frac{1}{8}[\tilde{\alpha}_2\log_2\tilde{\alpha}_2+\tilde{\alpha}_3\log_2\tilde{\alpha}_3\nonumber \\
&+&(\tilde{\beta}_+-\tilde{\gamma}_+)\log_2(\tilde{\beta}_+-\tilde{\gamma}_+)+(\tilde{\beta}_--\tilde{\gamma}_-)\log_2(\tilde{\beta}_--\tilde{\gamma}_-)]\nonumber \\
&-&\frac{1}{16}[(\tilde{\beta}_++\tilde{\gamma}_++\tilde{\alpha}_1+\sqrt{\tilde{\beta}_+^2+2\tilde{\beta}_+\tilde{\gamma}_++17\tilde{\gamma}_+^2-2\tilde{\beta}_+\tilde{\alpha}_1-2\tilde{\gamma}_+\tilde{\alpha}_1+\tilde{\alpha}_1^2})\nonumber \\
&&\log_2(\tilde{\beta}_++\tilde{\gamma}_++\tilde{\alpha}_1+\sqrt{\tilde{\beta}_+^2+2\tilde{\beta}_+\tilde{\gamma}_++17\tilde{\gamma}_+^2-2\tilde{\beta}_+\tilde{\alpha}_1-2\tilde{\gamma}_+\tilde{\alpha}_1+\tilde{\alpha}_1^2})\nonumber \\&+&(\tilde{\beta}_++\tilde{\gamma}_++\tilde{\alpha}_1-\sqrt{\tilde{\beta}_+^2+2\tilde{\beta}_+\tilde{\gamma}_++17\tilde{\gamma}_+^2-2\tilde{\beta}_+\tilde{\alpha}_1-2\tilde{\gamma}_+\tilde{\alpha}_1+\tilde{\alpha}_1^2})\nonumber \\
&&\log_2(\tilde{\beta}_++\tilde{\gamma}_++\tilde{\alpha}_1-\sqrt{\tilde{\beta}_+^2+2\tilde{\beta}_+\tilde{\gamma}_++17\tilde{\gamma}_+^2-2\tilde{\beta}_+\tilde{\alpha}_1-2\tilde{\gamma}_+\tilde{\alpha}_1+\tilde{\alpha}_1^2})
\\&+&(\tilde{\beta}_-+\tilde{\gamma}_-+\tilde{\alpha}_4+\sqrt{\tilde{\beta}_-^2+2\tilde{\beta}_-\tilde{\gamma}_-+17\tilde{\gamma}_-^2-2\tilde{\beta}_-\tilde{\alpha}_4-2\tilde{\gamma}_-\tilde{\alpha}_4+\tilde{\alpha}_4^2})\nonumber \\
&&\log_2(\tilde{\beta}_-+\tilde{\gamma}_-+\tilde{\alpha}_4+\sqrt{\tilde{\beta}_-^2+2\tilde{\beta}_-\tilde{\gamma}_-+17\tilde{\gamma}_-^2-2\tilde{\beta}_-\tilde{\alpha}_4-2\tilde{\gamma}_-\tilde{\alpha}_4+\tilde{\alpha}_4^2})\nonumber \\
&+&(\tilde{\beta}_-+\tilde{\gamma}_-+\tilde{\alpha}_4-\sqrt{\tilde{\beta}_-^2+2\tilde{\beta}_-\tilde{\gamma}_-+17\tilde{\gamma}_-^2-2\tilde{\beta}_-\tilde{\alpha}_4-2\tilde{\gamma}_-\tilde{\alpha}_4+\tilde{\alpha}_4^2})\nonumber \\
&&\log_2(\tilde{\beta}_-+\tilde{\gamma}_-+\tilde{\alpha}_4-\sqrt{\tilde{\beta}_-^2+2\tilde{\beta}_-\tilde{\gamma}_-+17\tilde{\gamma}_-^2-2\tilde{\beta}_-\tilde{\alpha}_4-2\tilde{\gamma}_-\tilde{\alpha}_4+\tilde{\alpha}_4^2})
],\hspace{0.7cm}
\end{eqnarray}
and
\begin{eqnarray}\label{en9}
S(\Pi(\rho_W^d))&=&3-\frac{1}{8}[\tilde{\alpha}_1\log_2\tilde{\alpha}_1+\tilde{\alpha}_2\log_2\tilde{\alpha}_2+\tilde{\alpha}_3\log_2\tilde{\alpha}_3+\tilde{\alpha}_4\log_2\tilde{\alpha}_4\nonumber \\
&+&(\tilde{\beta}_++\tilde{\gamma}_+)\log_2(\tilde{\beta}_++\tilde{\gamma}_+)+(\tilde{\beta}_+-\tilde{\gamma}_+)\log_2(\tilde{\beta}_+-\tilde{\gamma}_+)\nonumber \\
&+&(\tilde{\beta}_-+\tilde{\gamma}_-)\log_2(\tilde{\beta}_-+\tilde{\gamma}_-)+(\tilde{\beta}_--\tilde{\gamma}_-)\log_2(\tilde{\beta}_--\tilde{\gamma}_-)],
\end{eqnarray}
the quantum correlation reads
\begin{eqnarray}
{\cal M}(\rho_W^d)&=&-\frac{(1+e^{-8\kappa t})}{2}-\frac{1}{8}[\tilde{\alpha}_1\log_2\tilde{\alpha}_1+\tilde{\alpha}_4\log_2\tilde{\alpha}_4\nonumber \\
&+&(\tilde{\beta}_++\tilde{\gamma}_+)\log_2(\tilde{\beta}_++\tilde{\gamma}_+)+(\tilde{\beta}_-+\tilde{\gamma}_-)\log_2(\tilde{\beta}_-+\tilde{\gamma}_-)]\nonumber \\
&+&\frac{1}{16}[(\tilde{\beta}_++\tilde{\gamma}_++\tilde{\alpha}_1+\sqrt{\tilde{\beta}_+^2+2\tilde{\beta}_+\tilde{\gamma}_++17\tilde{\gamma}_+^2-2\tilde{\beta}_+\tilde{\alpha}_1-2\tilde{\gamma}_+\tilde{\alpha}_1+\tilde{\alpha}_1^2})\nonumber \\
&&\log_2(\tilde{\beta}_++\tilde{\gamma}_++\tilde{\alpha}_1+\sqrt{\tilde{\beta}_+^2+2\tilde{\beta}_+\tilde{\gamma}_++17\tilde{\gamma}_+^2-2\tilde{\beta}_+\tilde{\alpha}_1-2\tilde{\gamma}_+\tilde{\alpha}_1+\tilde{\alpha}_1^2})\nonumber \\&+&(\tilde{\beta}_++\tilde{\gamma}_++\tilde{\alpha}_1-\sqrt{\tilde{\beta}_+^2+2\tilde{\beta}_+\tilde{\gamma}_++17\tilde{\gamma}_+^2-2\tilde{\beta}_+\tilde{\alpha}_1-2\tilde{\gamma}_+\tilde{\alpha}_1+\tilde{\alpha}_1^2})\nonumber \\
&&\log_2(\tilde{\beta}_++\tilde{\gamma}_++\tilde{\alpha}_1-\sqrt{\tilde{\beta}_+^2+2\tilde{\beta}_+\tilde{\gamma}_++17\tilde{\gamma}_+^2-2\tilde{\beta}_+\tilde{\alpha}_1-2\tilde{\gamma}_+\tilde{\alpha}_1+\tilde{\alpha}_1^2})
\\&+&(\tilde{\beta}_-+\tilde{\gamma}_-+\tilde{\alpha}_4+\sqrt{\tilde{\beta}_-^2+2\tilde{\beta}_-\tilde{\gamma}_-+17\tilde{\gamma}_-^2-2\tilde{\beta}_-\tilde{\alpha}_4-2\tilde{\gamma}_-\tilde{\alpha}_4+\tilde{\alpha}_4^2})\nonumber \\
&&\log_2(\tilde{\beta}_-+\tilde{\gamma}_-+\tilde{\alpha}_4+\sqrt{\tilde{\beta}_-^2+2\tilde{\beta}_-\tilde{\gamma}_-+17\tilde{\gamma}_-^2-2\tilde{\beta}_-\tilde{\alpha}_4-2\tilde{\gamma}_-\tilde{\alpha}_4+\tilde{\alpha}_4^2})\nonumber \\
&+&(\tilde{\beta}_-+\tilde{\gamma}_-+\tilde{\alpha}_4-\sqrt{\tilde{\beta}_-^2+2\tilde{\beta}_-\tilde{\gamma}_-+17\tilde{\gamma}_-^2-2\tilde{\beta}_-\tilde{\alpha}_4-2\tilde{\gamma}_-\tilde{\alpha}_4+\tilde{\alpha}_4^2})\nonumber \\
&&\log_2(\tilde{\beta}_-+\tilde{\gamma}_-+\tilde{\alpha}_4-\sqrt{\tilde{\beta}_-^2+2\tilde{\beta}_-\tilde{\gamma}_-+17\tilde{\gamma}_-^2-2\tilde{\beta}_-\tilde{\alpha}_4-2\tilde{\gamma}_-\tilde{\alpha}_4+\tilde{\alpha}_4^2})
].
\hspace{1cm}
\end{eqnarray}

Quantum correlation obtained by AMID for this  case is ${\cal
A}(\theta_1,\phi_1,0,\theta_2,\phi_2,0,\theta_3,\phi_3,0)$ coincides
with the one obtained by MID (black line in Fig.~\ref{fig2}). The
MID and AMID for the three-qubit initial W state under various noisy
channels are depicted in Fig.~\ref{fig2}. As it can be seen from the
figure, the quantum correlations obtained by MID for two cases of
Pauli-X and -Y channels are overestimated with respect to ones
obtained by AMID. Notice that, for the initial W state our results
do not agree with the results obtained by quantum discord which is
due to the different choices of the projective measurements
\cite{mah12}.

\section{Conclusions}\label{sec5}
In this paper, we have studied quantum correlations for the initial
GHZ and W states in the presence of various noisy channels using the
measurement-induced disturbance and its ameliorated version. We
considered the solutions of the Lindblad equation where the noises
are represented by the Pauli-X, Pauli-Y, Pauli-Z and isotropic
operators. This idea is based on the fact that the classical
measurements can be performed without disturbance. However,
measurements usually disturb the system in the quantum description
and this disturbance can be used to determine the quantumness of
correlations. In the absence of noise, quantum correlations of GHZ
and W states are equal to unity, namely the half of the total
correlation which is expected for a pure state. After turning on
noises, quantum correlation decreases for all noises. For the case
of the initial W state under Pauli-Y noisy channel (unlike GHZ state
that its corresponding quantum correlation vanishes for large
$\kappa t$), ${\cal A}(\rho_{W}^y)$ tends to $0.58$ as $\kappa t$
goes to infinity. In comparison, our results showed that in the MID
approach the W state is more robust than GHZ state under noisy
channels except Pauli-X channel. This result is also valid for the
AMID scenario except Pauli-Y channel for $0<\kappa t<0.4$. Moreover,
the obtained results for ${\cal M}(\rho)$ coincided with those of
quantum discord just for the initial GHZ state. Indeed, both the
quantum discord and MID overestimate quantum correlations of states
with respect to AMID in agreement with Ref.~\cite{girolami11}.

\acknowledgments We would like to thank Robabeh Rahimi for fruitful
discussions and suggestions and for a critical reading of the paper.

\end{document}